\documentclass{an}
\usepackage{graphicx}
\usepackage{times}
\usepackage{fancyhdr}
\begin{document}

\title{V4332 Sgr in 'Quiescence'\thanks{Based on observations collected at the
European Southern Observatory, La Silla and on observations
collected at Asiago observatory}}

\author{S. Kimeswenger}
\institute{Institut f\"ur Astrophysik der Leopold--Franzens--Universit\"at
Innsbruck, Technikerstr. 25, A-6020 Innsbruck, Austria}

\date{Received; accepted; published online}

\abstract{ In 1994 V4332 Sgr underwent a mysterious eruption.
Somehow its fast evolution towards a red giant star was, lacking
alternative classifications, connected to the red variable M31 RV,
which had its eruption in 1988. The red eruptive variable V838~Mon
draw in February 2002 the attention back to its 'older twin'
V4332~Sgr. The new precise photometry of the progenitor given here
shows that the object started to rise years before the 1994 event.
Post outburst photometry and spectroscopy from 2002 and 2003 show
that the object stopped it's decline and seem to reheat now. The
progenitor data and the new high quality spectra provide a
supplement and completion to the data around the outburst given by
Martini et al. (1999). It thus allows theorists to give new
boundaries for modelling of this unusual object. }

\correspondence{Stefan.Kimeswenger@uibk.ac.at}

\maketitle


%

\section{Introduction}

V4332 Sgr (Nova Sgr 1994 1) was discovered on February 24, 1994 by
Hayashi et al. \cite{discovery}. Already first spectra (Wagner
1994) indicated that this object was not a classical nova early in
outburst. It lacked the spectral features characteristic for this
class of objects. It then changed its spectral type from
approximately M0 to M5 in 5 days only (Tomaney et al. 1994). Based
on this kind of evolution they concluded that V4332 Sgr underwent
a similar eruption like the luminous red variable in the bulge of
M31 (M31 RV) discovered by Rich et al. \cite{rich89}. A detailed
review of the photometric and spectroscopic evolution of V4332~Sgr
during the decline is given by Martini et al. \cite{martini99}. To
my knowledge, the last spectra before those presented here were
obtained June 5-6, 1994 when the star had dropped to $V=18\fm7$.
The same seems to apply to the photometry - except those from
survey archives discussed later in this paper and a narrow band
H$\alpha$ image obtained with HST in 1997.

I present here own spectroscopic observations and photometry
obtained in August 2002 and in July 2003. Careful searches through
the archives provided some additional data from HST (H$\alpha$
image taken November 3$^{\rm rd}$, 1997) and from the Asiago
Observatory ($VR_{\rm C}I_{\rm C}$ frames taken May 16$^{\rm th}$,
2002). Finally I carefully collected and calibrated the progenitor
data to have a better data base than that given in the first
estimates by Wagner et al. (1994). As only the 1985 to 1992 sky
survey plates are available in this region as digitized sky survey
(DSS), I digitized with a CCD camera and a microscope POSS-I plate
copies (observed 1950.53).

These data are discussed in connection with other data: low S/N
optical spectra obtained in April 2003 (Tylenda et al. 2004) and
in September 2003 (Banerjee \& Ashok 2004) and near infrared data
from DENIS, 2MASS and Banerjee et al. (2003, 2004a,b). Although
there is an observational gap after the 1994 outburst it may
provide clues to construct an unambiguous picture of this object.
The spectra contain numerous strong unidentified emission
features. These should be subject of a NLTE moving model
atmosphere - including molecules.

\section{Photometry}

\subsection{The interstellar extinction}
Using the spectroscopic classification of Martini et al.
\cite{martini99} and the photometry of Gilmore et al. (1994) and
Wagner et al. (1994) for March 4 and 9 1994, I obtain an
\mbox{E$_{\rm V-I}\approx 0\fm59$} \mbox{($\Rightarrow$ E$_{\rm
B-V}\approx 0\fm42$)}, E$_{\rm V-R}\approx 0\fm41$
\mbox{($\Rightarrow$ E$_{\rm B-V}\approx 0\fm35$)}, and E$_{\rm
B-V}\approx 0\fm29$. Although the star had changed from K4 to M3
the extinction values are consistent within the range of same
bands for the two dates within 0\fm01. As the red bands are less
affected by stellar lines and thus possible abundance effects, I
give them a higher weight and thus get E$_{\rm B-V}\approx
0\fm37\pm0.07$ for V4332 Sgr. This is slightly higher but within
the errors consistent with the value of 0\fm32 given by Martini et
al. \cite{martini99}.

\begin{figure}
\centerline{\resizebox{\columnwidth}{!}{\includegraphics{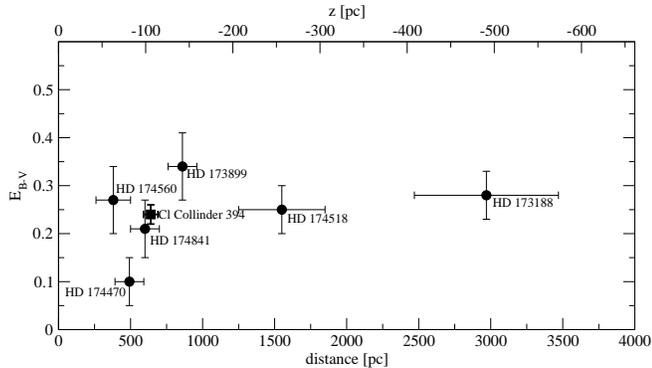}}}
 \caption{The extinction-distance diagram for stars with well known spectral
 class and luminosity within 45\arcmin  around V4332 Sgr. Additionally
 the well studied open cluster Collinder 394 is given. The extinction
 is constant for distances of $>$ 500 pc ($\equiv$ z $\approx$ 100 pc;
 upper axis).
} \label{ebv_r}
\end{figure}

From the SIMBAD data base all stars within a projected distance of
45\arcmin\ from V4332 Sgr having well defined photometry and
spectral types and a stellar open cluster were picked to derive an
extinction-distance diagram. Recently Fr\"obich et al. (2005) used
2MASS star counts to derive the differential extinction for the
whole galactic plane. It shows no variation within a square degree
around the target. This justifies the use of such a large field.
The results (Figure \ref{ebv_r}) show nearly no additional
extinction left after a distance of 500\ pc. The result also
corresponds well to the total extinction of the Galaxy given for
this field by Schlegel et al. (1998). Thus the extinction cannot
be used to derive a distance like it is often used for such kind
of unique objects (e.g. for V4334 Sgr by Kimeswenger (2002) or for
V838 Mon by Munari et al. (2005))

\subsection{Photometry of the progenitor}

\begin{figure}
%
{\noindent\resizebox{\columnwidth}{!}{\includegraphics{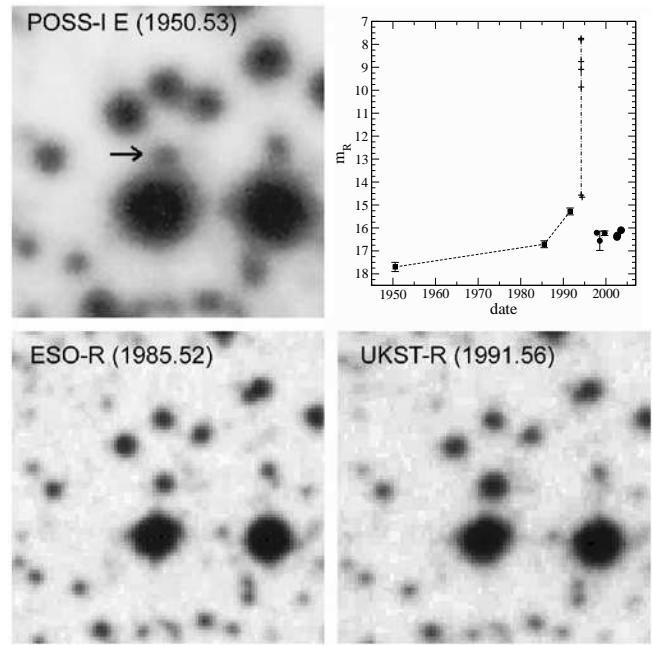}}}

\vspace{-8.4cm}
{\noindent\phantom{x}\hspace{4.2cm}\resizebox{4cm}{!}{\includegraphics{F2B.eps}}}
\vspace{4.4cm}

\caption{The pre--outburst sky survey plates in the red band. The strong
brightening is obvious here already. The red band photometry (insert) shows the
total evolution of the target including the outburst data from Martini et al.
(1999) and the post-outburst photometry derived here.} \label{surveys}
\end{figure}

Sky survey plates scans from the pre-outburst period are available
from the various online archives (see Figure \ref{surveys}). As
the POSS-I plates are not available as scans south of $\delta =
-17\degr$ I digitized with a CCD camera and a microscope POSS-I
plate copies available at my institute.  Further searches in
plates archives of telescopes like Tautenburg, Kiso, Asiago etc.
were carried out. But I failed to get additional pre-outburst
plates being deep enough to show the target. I calibrated the
scans using differential photometry and my CCD sequence in the
field. {For the conversion of the CCD magnitudes to the plate
scale and the error estimates the recently published method my
Bacher et al. (\cite{dss}) was used. As shown there the error
estimates are conservative.} The Quick-V (1987.59) survey is not
deep enough to give a reliable photometry of the target.

\begin{table}
\caption{Photometry of the pre-outburst phase.}
\begin{tabular}{l c c c}
plate & date & band & mag \\
\hline
POSS-I blue & 1950.53 & $O$ & $19\fm23\pm0\fm12$\\
POSS-I red & 1950.53 & $E$ & $17\fm72\pm0\fm10$\\
UKST-B & 1976.57 & $B_J$ & $18\fm01\pm0\fm07$ \\
ESO-R & 1985.52 & $R_{59}$ & $16\fm72\pm0\fm05$\\
UKST-R & 1991.59 & $R_{61}$ & $15\fm28\pm0\fm05$\\
\end{tabular}
\end{table}

The star brightened in the red band ($E$ and $F$ bands are nearly
identical to $R$ and thus comparable; H\"ortnagl et al. 1992) by a
factor of 10. This remarkable change cannot originate from a
eclipsing binary system of normal main sequence stars as it is
suggested in the model of Tylenda et al. (2004). The colors of
this variable star from multi--epoch data (e.g. in the SuperCOSMOS
catalogue) should not be used to estimate a spectral type and thus
e.g. a distance. The only single epoch - multi color data point is
available with POSS-I surveys (the plates were taken within the
same night). With the interstellar reddening derived above, and
the POSS-I photometry (using the color equations of Dorschner et
al. 1966) I derive a spectral type of G2 ($\pm$ 0.4 subclasses).
This results in a distance of 2.9, 5.8 and 10 kpc ($\approx$ 30\%
uncertainty) for luminosity class {\tt V}, {\tt IV} and {\tt III}
respectively. Thus the distance of 300 pc derived by Martini et
al. (1999) is very unlikely - it should have been an extremely
subluminous object. Such distances also fit well to the
extinction-distance diagram given above. Tylenda et al. (2004)
using
 $B_J$ from 1976 and $R$ from 1985 (both from the SuperCOSMOS
catalogue) get, due to the steady rise in all bands, a later spectral type (G6
to K0) and thus a more nearby system.

Using the color equation of Blair \& Gilmore (1982) for the $B_J$
band I derive a rise in $B$ band from 1950 to 1976 by 0\fm9 to
1\fm0 (uncertain by about 0\fm2 due to the color equation). As
this is about the same amount like the total red increase from
1950 to 1985, one have to assume that the object got redder
($\Delta_{B-V} \approx 0\fm2 - 0\fm4$). Thus the bolometric
luminosity rises a little bit less than the factor of 10 found in
the $R$ magnitude.

{The only epoch were 2 bands were taken more or less
simultaneously was 1950. Thus color estimates and thus estimates
of the spectral type of the progenitor are extremely difficult.
Color equations derived for normal stars were applied. Thus this
determination is only reliable under the assumption that the
object was not an emission line star at that time.}

\subsection{The post-outburst photometry}

I have taken direct images with $VR_{\rm C}I_{\rm C}$ bands
shortly after the spectra discussed later. Additionally Martini
\cite{Ma03} provided me with images taken a few weeks after my
2002 ESO run at the Las Campanas 2.5m telescope. Both - absolute
and differential photometry was carried out with these images. The
original DeNIS survey image ($I_{\rm C}$) was taken from the
consortium data base and calibrated differentially with the NTT
CCD frames. This gives a better result than the survey
calibration. In the Asiago Observatory archive I found a set of
direct images taken May 2002. The accuracy of the photometry given
in Table~\ref{phot_tab} is better than $0\fm03$ throughout all
bands.

\begin{table}
\caption{Photometric results from CCD direct imaging:}
\label{phot_tab}
\begin{tabular}{lllcc}
date & UT & telescope & band & mag\\
\hline
11. Sep. 1999 & 00:29 &  DeNIS      & $I_{\rm C}$ & 14\fm37 \\
16. May. 2002 & 02:35 &  Asiago     & $V$         & 17\fm76 \\
              &          &          & $R_{\rm C}$ & 16\fm37 \\
              &          &          & $I_{\rm C}$ & 14\fm75 \\
~6. Aug. 2002 & 00:30 &  ESO NTT    & $V$         & 17\fm74 \\
              &          &          & $R_{\rm C}$ & 16\fm39 \\
              &          &          & $I_{\rm C}$ & 14\fm84 \\
31. Aug. 2002 & 23:50 &  LCO-100    & $V$         & 17\fm77 \\
              &          &          & $R_{\rm C}$ & 16\fm33 \\
              &          &          & $I_{\rm C}$ & 14\fm92 \\
20. Jul. 2003 & 01:20 &  ESO NTT    & $V$         & 17\fm59 \\
              &          &          & $R_{\rm C}$ & 16\fm10 \\
\smallskip
              &          &          & $I_{\rm C}$ & 14\fm95 \\
\hline
\end{tabular}
\end{table}

\begin{figure}
\centerline{\resizebox{\columnwidth}{!}{\includegraphics{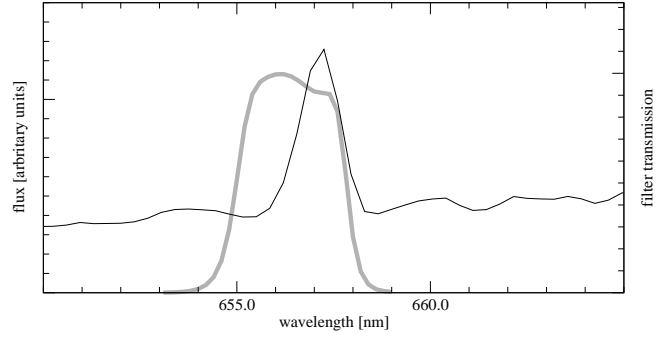}}}
 \caption{The CaI\,\,(657.3nm) contamination in the HST H$\alpha$ filter (grey line).
} \label{hst}
\end{figure}

\begin{figure}
\centerline{\resizebox{\columnwidth}{!}{\includegraphics{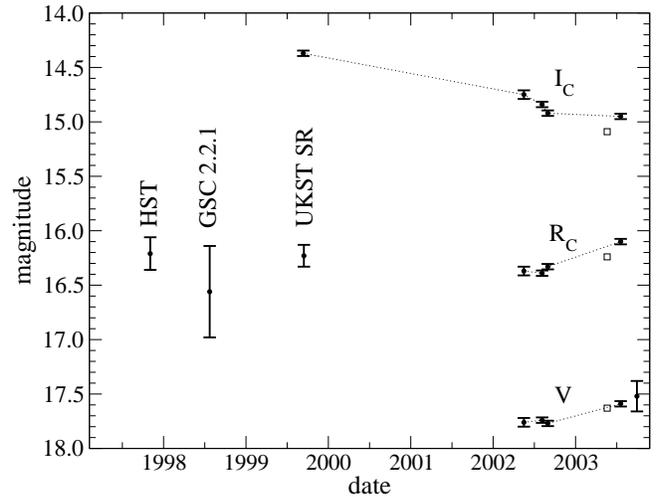}
}} \caption{Recent photometric evolution of V4332 Sgr. While the
$I_{\rm C}$ flux faded 1999 to 2002 it was stable thereafter. The
flux in the $V$ and $R$ bands started to rise again.}
\label{phot_fig}
\end{figure}

In the HST archive a H$\alpha$ image taken November 3$^{\rm rd}$,
1997 was found. This filter is well centered on the effective
wavelength of the $R$ band. As the CaI\,\,(657.3nm) line
contributes to the flux, I folded the spectrum with the filter
curve. This was used to correct for the contamination by this
emission line (Figure \ref{hst}). To test the procedure the 2002
and 2003 $R$ band photometry was compared in the same way by using
"artificial" images folding the spectra with the published filter
curve. There the deviation was less than 0\fm05. The error using
the UKST-H$_{\alpha}$ photographic survey (taken 2002.5) in the
same way is 0\fm1. Thus the error estimate of 0\fm15 given in
Figure \ref{phot_fig} is certainly conservative.

Banerjee \& Ashok \cite{ban04} published photometry from 29$^{\rm
th}$ Sep. 2003. I included the $V$ band in the analysis. For the
$R$ and $I$ band it is not clear whether Cousins, Bessel or
Johnson filters were used. The GSC 2.2.1 gives us $R = 16\fm 56$
for 1998.54. Comparing about 10 nearby stars from that GSC
calibration with my CCD set leads to a shift of +0\fm02 with a rms
of 0\fm07. Thus the error of 0\fm42 given in the catalogue is, for
my point of view, too high. Finally a short 4 minute exposure red
plate, taken as continuum subtraction image for the
UKST-H$_{\alpha}$ survey, from 1999.7 exists. It was calibrated in
the same way as the progenitor photometry.

As shown in Figure~\ref{phot_fig} the target stopped it's decline and started
to get even a little bit hotter now. However this interpretation of the broad
band photometry has to be used carefully. The spectrum is dominated by emission
lines, producing a significant fraction of the flux. Banerjee et al.
\cite{ban03} find a decline in $J$ between the 1998 2MASS measurements and
their 2003 data. Taking the ($R-J$), which changed from $\approx 4\fm0$ in 1998
to 2\fm89 in 2003, gives another signature of heating -- on the other hand the
$K$ band shows an increasing IR excess. Thus the bolometric luminosity clearly
increased from 1998 to 2003.

\begin{figure}
\centerline{\resizebox{\columnwidth}{!}{\includegraphics{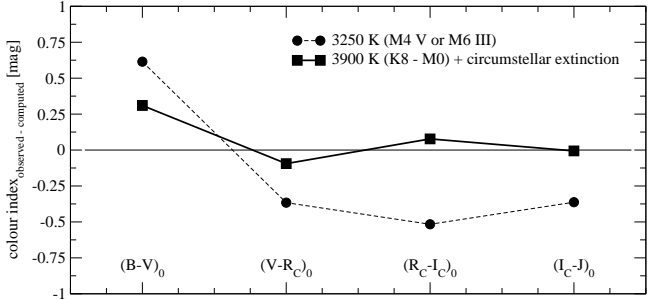}}}
 \caption{The temperature of the "photosphere" during 2003. The K8-M0 star
 adding an additional circumstellar extinction gives a much better fit
 than the M4-M6 star derived by Banerjee et al. (2003).
} \label{color_index}
\end{figure}

Assuming the contamination by the dust shell built during the late
90s to be negligible in the $J$ band (Banerjee et al. 2003) I
derived, using intrinsic colors of real stars (Cox 2000), an
effective temperature of 3.900\,K and a circumstellar extinction,
additional to the interstellar extinction derived during the
outburst phase, of E$_{B-V}$=0\fm44$\pm$0\fm05. This temperature
is significantly higher than the 3.250\,K derived by the simple
blackbody fit in Banerjee et al. \cite{ban03}. As shown in Figure
\ref{color_index} the hotter star with the additional extinction
is consistent over a large wavelength range. This results in a
temperature of 865$\pm$20\,K for the temperature of the dust shell
causing the near infrared excess. This is slightly lower than the
previously derived temperature of 900\,K (Banerjee et al. 2003).

\section{Astrometry}
To possibly estimate a proper motion (if the distance of 300\,pc of Martini et
al. 1999 applies), the images and the four sky survey plates were calibrated
astrometrically. For the calibration in the modern ICRS 2000 frame, I used the
{\it USNO CCD Astrometric Catalogue} ({\tt UCAC}) (Zacharias et al. 2000).
Andersen \& Kimeswenger \cite{nova01} show in detail the accuracy of this
method and strict bounding to the {\tt TYCHO-2} reference frame. This leads to
the very accurate coordinates on the CCD frames (epoch 2002.66 / equinox
J2000.0)

\medskip
\begin{tabular}{lcll}
$\alpha_{\rm ICRS}$ & = & 18$^{\rm h}$50$^{\rm m}$36\fs713 & $\pm$ 0\fs005\\
$\delta_{\rm ICRS}$ & = &-21\degr23\arcmin28\farcs94\ & $\pm$ 0\farcs03\\
\end{tabular}
\medskip

The numerical solution for the proper motion for the sky survey
plates, with the epochs 1976.567 (UKST-J), 1985.519 (ESO-R),
1986.686 (UKST-IR) and 1991.581 (UKST-R) gives no significant
proper motion. The HST archive image (H$\alpha$ narrow band) taken
1997.841 covers only a very small region around the target. It
thus was calibrated relative to the global solution obtained
above. Using 8 surrounding stars relative to the 2002.66 and the
2003.61 NTT images gives an upper limit of 4~mas/yr. As we are
looking towards the galactic center, differential shear due to the
rotation should be significant. This upper limit, not necessarily
but most likely, excludes also a short distance scale of 300\,pc
of Martini et al. \cite{martini99}.

\begin{figure*}
\centerline{\includegraphics[width=175mm]{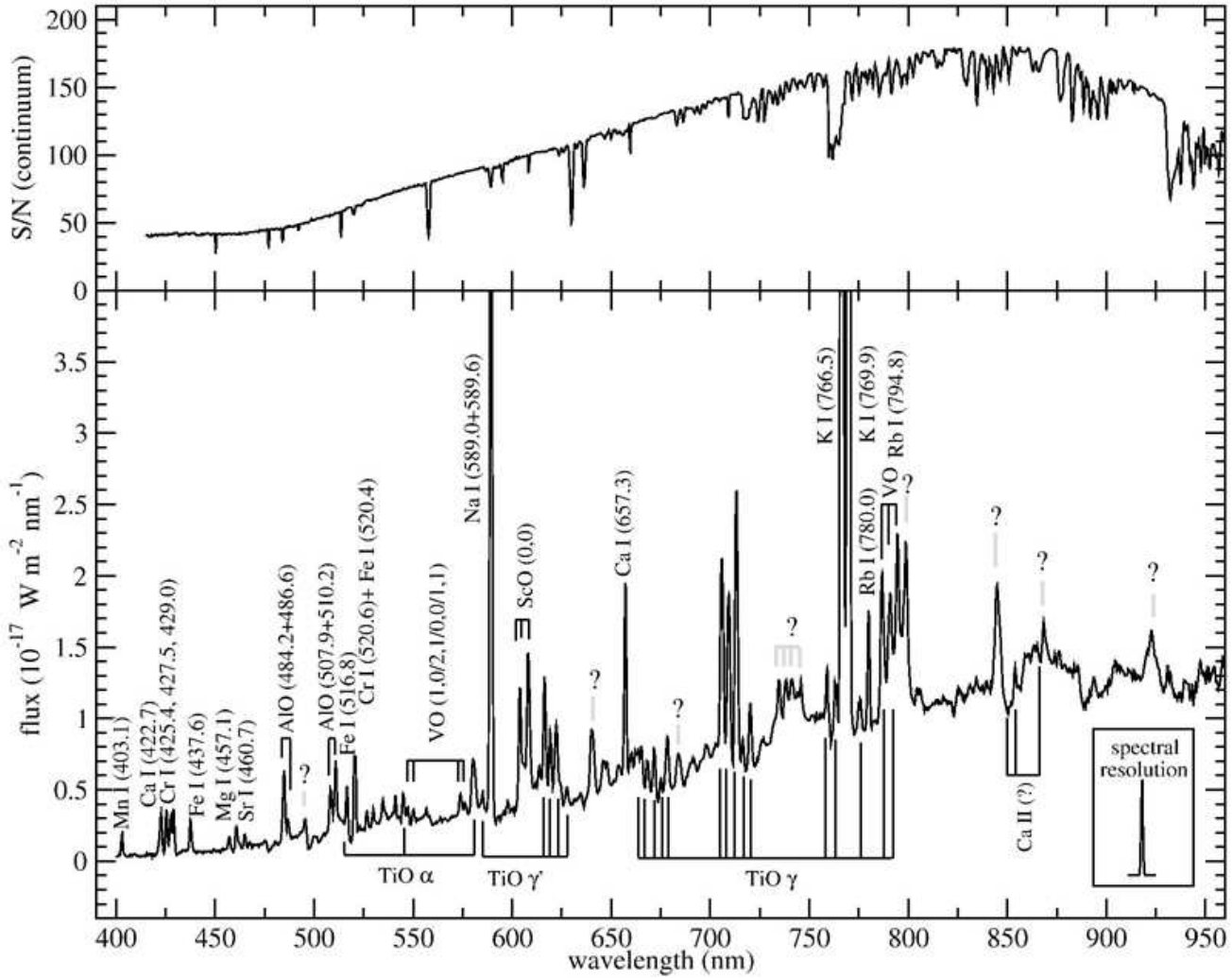}} \caption{The
spectrum of V4332 Sgr taken in July 2003. The NaI (589.0 + 589.6
nm) and the KI (766.5 and 769.9nm) are truncated in order to see
all the weaker features. The slope of the continuum is consistent
with the photometric results. The main features are marked.
Strong, but unidentified features are marked with grey tics. {The
insert shows the gaussian with the FWHM of the night sky lines.
The upper panel shows the calculated S/N ratio from the data
reduction (continuum). In the region of the telluric bands the
deviation between the standard stars was taken and multiplied with
a factor of 3 to achieve a conservative estimate for the errors.}
} \label{full_spectrum}
\end{figure*}

\section{Spectroscopy}
\subsection{Data}
The spectra were taken at the ESO New Technology Telescope August
6, 2002 (+EMMI red arm) and July 20--23, 2003 (+EMMI red \& blue
arm). At the red arm grism \#2 and slit width of 1\farcs0 was
used. This gives a resolution (binning the CCD 2$\times$2) of
$\approx$0.3nm/pixel. The FWHM of night sky lines is 0.91nm,
giving us the real resolution. The usable range of the spectrum
spans from 450nm to 960nm. In July 2003 spectra with the EMMI blue
arm and grating \#5 were taken, resulting in a similar resolution
and in a range of 350nm to 550nm. As the signal was very weak
below 400nm this region was not used for the final spectrum. The
data were reduced, using standard techniques in {\tt MIDAS} and
calibrated using the standard stars G93-48, LT~6248 and EG~274.
{The full observational log is given in Table~\ref{spec_log}.} The
2002 spectrum (medium S/N) and 2003 spectra (high S/N) were
overlayed. Except the changes in the KI and in the NaI lines (see
below), no significant variations (within the errors) were found.
Thus I was able to combine them. The spectrum is shown in
Figure~\ref{full_spectrum}. The spectrum may be obtained from the
author electronically on request. All the tiny structures down to
$5\,\times\,10^{-19}$ W m$^{-2}$ nm$^{-1}$ (half a tick in
Figure~\ref{full_spectrum}) are visible in all spectra and thus
have to be real. {The S/N was derived from the MIDAS data
reduction for the continuum for a single spectrum. Additionally
deviations in the region of the telluric bands was taken from the
standard stars and multiplied by a factor of 3 to achieve a
conservative error estimate. The S/N in the strong lines thus is
even much better. Furthermore the combination of all spectra gives
another improvement of about a factor of 2. Another check for the
reliability is derived by dividing the spectrum obtained 2003 July
20 and that obtained July 23. Even at the position of the telluric
bands the deviations are below 3\%.}

\begin{table}[ht]
\caption{Observational log of the spectroscopy:} \label{spec_log}
\begin{tabular}{llrcc}
date & UT & exp. & mode & airmass  \\
 & start & time     &  &  seeing\\
\hline
~6. Aug. 2002 & 00:19 &  300  & EMMI Red Arm  & 1.17   \\
              &       &       & Grism \#2   & 1\farcs3 \\
              & 00:25 &  300  & EMMI Red Arm  & 1.16   \\
              &       &       & Grism \#2   & 1\farcs3 \\
20. July 2003 & 00:37 & 1200  & EMMI Red Arm  & 1.36   \\
              &       &       & Grism \#2     & 1\farcs4 \\
              & 01:20 &  250  & EMMI Blue Arm & 1.19   \\
              &       &       & Grating \#5   & 1\farcs5 \\
              & 01:30 & 1800  & EMMI Blue Arm & 1.17   \\
              &       &       & Grating \#5   & 1\farcs3 \\
21. July 2003 & 23:35 & 1800  & EMMI Blue Arm & 1.71   \\
              &       &       & Grating \#5   & 0\farcs9 \\
22. July 2003 & 00:15 & 1200  & EMMI Red Arm  & 1.42   \\
              &       &       & Grism \#2     & 0\farcs8 \\
23. July 2003 & 00:05 & 1800  & EMMI Blue Arm & 1.46   \\
              &       &       & Grating \#5   & 1\farcs0 \\
              & 00:43 & 1200  & EMMI Red Arm  & 1.28   \\
\smallskip
              &       &       & Grism \#2   & 1\farcs1 \\
\hline
\end{tabular}
\end{table}

\subsection{Line identification}
As already pointed out by Banerjee \& Ashok \cite{ban04} the
spectrum is dominated by the NaI (589.0 + 589.6\,nm) and the KI
(766.5 and 769.9\,nm) lines. The S/N of my composite spectrum is
significantly better than that given there. Thus it is easier to
distinguish between bundles of blended lines and continuum. This
is essential to define the real fluxes (resp. equivalent widths)
of the lines. The identifications of the strong features found
by Banerjee \& Ashok \cite{ban04} mostly can be confirmed.\\
Their identification of VO(0,0) 608.6\,nm is more likely
connected, together with the strong feature around 603.5\,nm, to
ScO(0,0) (603.6 + 608.0\,nm) - especially as overtones fill the
gap between them in the same way as in VY\,\,CMa (Herbig 1974,
Wallerstein \& Gonz\'alez 2001). Their identification of TiO
$\gamma'$ (0,1) at 656.9\,nm should be more likely replaced by CaI
657.3\,nm ($\chi = 1.9$\,eV) due to missing other lines of this
TiO band. This CaI line also has an excitation very similar to the
strong KI and NaI features. Also the other line from the ground
state of CaI at 422.7\,nm (2.9eV) is very prominent.

 As strong low excitation metal lines (NaI and KI) changed while molecule
emission did not vary (see below), the classification of the 780.0
and 793.9nm features as RbI by Banerjee \& Ashok might be under
discussion. It should follow a change in line optical depth of the
emitting gas. These features might originate from the
TiO$\gamma$(2,3) and (3,4) series and the VO features as found
also in CY~CMa (Wallerstein 1971, Herbig 1974, Wallerstein \&
Gonz\'alez 2001) and in case of the unusual M giant U~Equ
(Barnbaum et al. 1996) as well. As the line ratio of these two
emissions cannot be reached by RbI for gas temperatures below
15.000K, at least some contamination has to exist.  Several other
strong features cannot be identified at all. Although it is nearly
always possible to find a low excitation line for a given
wavelength, as other lines of the same multiplet are missing these
identifications cannot be taken into account. The FeI
identifications have to be discussed. The ratio of the line at
516.8nm to that at 520.4\,nm (contaminated by CrI 520.6\,nm) and
the missing 522.5 and 626.9\,nm lines do not fit well to the
interpretation below. No lines with upper levels above
$\approx\,\,$3.1\,eV are found. This results in a electron
temperature for the emitting region of $\le$\,2.200\,K. {Table
\ref{spec_lines} gives all well defined lines and features. To
derive equivalent widths a Kurucz (1991) model having 3.900\,K was
used as continuum (see next section). This certainly is the
largest source of uncertainty in the line strengths.}

\begin{figure}[h]
\centerline{\includegraphics[width=85mm]{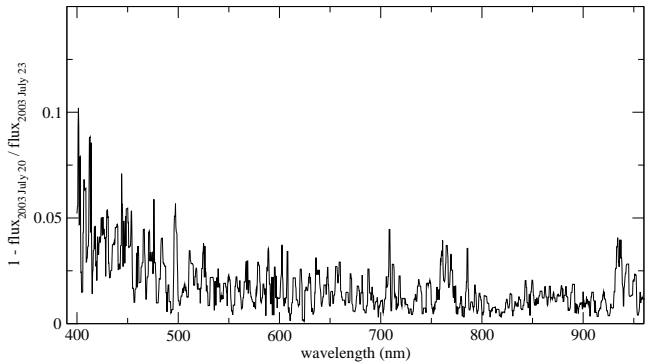}} \caption{The
quotient of the 2003 July 20 spectrum over the 2003 July 23
spectrum shows deviations of a few percent even in the region of
the telluric bands.} \label{dev_spectrum}
\end{figure}

\begin{figure*}[t]
\centerline{\includegraphics[width=160mm]{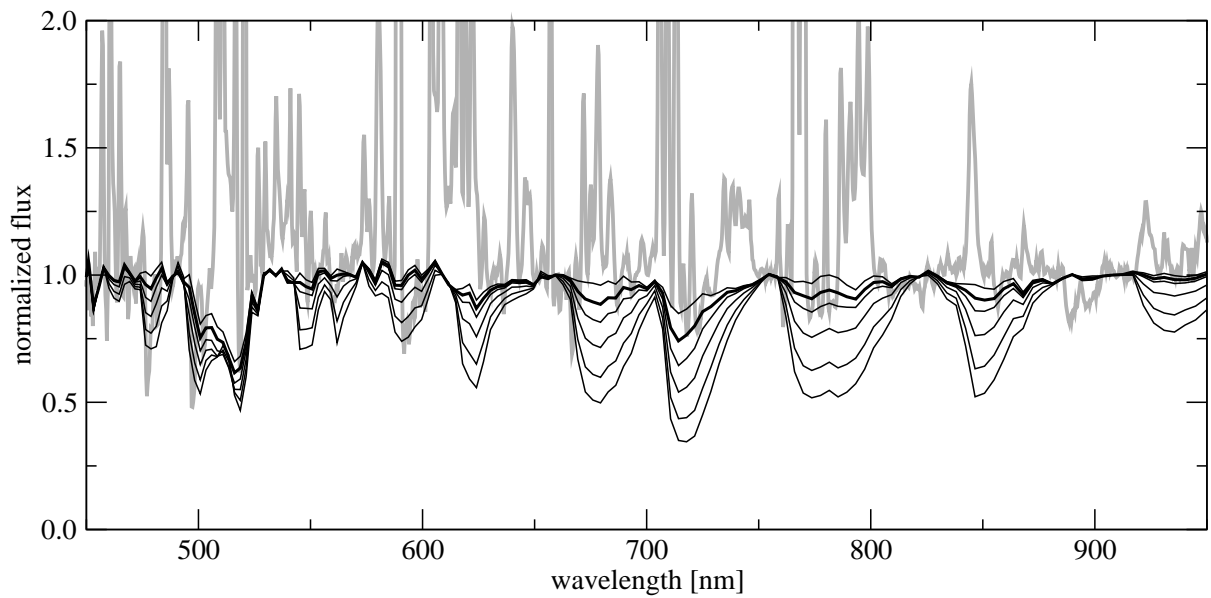}} \caption{The
normalized spectrum (grey line) and Kurucz (1991) model
atmospheres for 3.500 to 4.000\,K (step 100K).}
\label{cont_spectrum}
\end{figure*}

\begin{table*}[ht]
\caption{Lines of the spectrum: $\lambda_{obs} =\,\,$ observed
wavelength using gaussian fit;  $EW := \sum{(F_c - F)/ F_c\,\,}
d\lambda$} \label{spec_lines}
\include{lines_tab}
\end{table*}

\subsection{The continuum}
To derive, independently from the broad band photometry above, the
underlying continuum of the central source, a spectroscopic
comparison was started. To normalize the continuum only the lowest
points at 450, 530, 660, 705, 825 and 925\,nm were taken for a
spline fit. This restriction guaranties no contamination by
molecular bands even for very late spectral types. The resulting
spectrum was overlayed with Kurucz (1991) model atmospheres and
$\chi^2$ was derived. Most of the prominent absorption bands fit
very well with the 3.900\,K derived also in the photometry. Some
small bands may give a little bit lower temperature down to
3.700\,K - but none goes as far down as found by Banerjee et al.
\cite{ban03}. Thus the 3.900\,K photosphere of the central source
was adopted for the further analysis.

\subsection{Interpretation of the spectrum}

The molecular lines stay remarkably stable - both, between my
spectra in 2002 and 2003 as well as compared to the spectrum of
Banerjee \& Ashok \cite{ban04}. Only the 708.7\,nm
TiO$\gamma$(0-0) feature looks different in the spectrum of
Tylenda et al. (2004). I obtain a temperature of 600\,K for the
ScO lines using the calibration by Herbig (1974) and about 700\,K
for the TiO$\gamma$ lines using the transition strengths in the
online Kurucz tables. Banerjee et al. \cite{ban03} derived
somewhat higher values of 3.000\,K for the near infrared AlO
vibrational excitation. On the other hand they get only 300K for
the rotational terms. This emission either originates from
different spatial domains or from a region without thermodynamic
equilibrium. Generally the molecular lines fit well to the dust
temperature of 865\,K derived by the NIR excess.

The atomic lines unveil a very complex situation. The 766.5\,nm KI
line is affected by the telluric band. Thus it is critical to
derive the underlying continuum and thus the correct ratio of the
two lines. The total strength increased by 13\%~ between summer
2002 and summer 2003. The accuracy of about $\pm$5\% is due to the
strength of the lines in my spectra despite the problems with the
telluric bands. The ratio of the 766.5\,nm to the 769.9\,nm line
do not satisfy the expected one for (in the line) optically thin
media, (except if we go down to 147\,K - but then the emissivity
itself is negligible and the line should not be visible). Banerjee
\& Ashok \cite{ban04} use an access given by Williams
\cite{wil1994} to derive an optical depth for self absorption in
the line, assuming a ratio of 2:1 (which corresponds to an
electron temperature of $\ge 2.500$\,K). This access also assumes
an isothermal and homogenous plane-parallel slab. Remarkable is
that the optical depth decreased from $\tau = 4.9$ in 2002 to
$\tau = 3.4$ in 2003. Banerjee \& Ashok \cite{ban04} give for
September 2003 a value of 4.5. On the other hand the red line
(unaffected) increased further by another 10\%~ from July to end
of September 2003. Although this may be reached also by an
increase of the electron temperature, it is likely a decrease of
the optical depth. Thus it is unclear whether this turnaround is
real or just an effect of an improper continuum subtraction and
correction for the telluric band for the blue line in the low-S/N
spectra by Banerjee \& Ashok \cite{ban04}. The absolute values of
$\tau$ should not be used to derive column densities, as the less
optically thick line comes from a smaller but hotter region deeper
inside the circumstellar environment.  The increase of the line
strength in the NaI lines is even stronger (+27\%). Also here the
line increased further by another 16\%~ until September. As this
line is, due to the higher abundance of the element, much more
optically thick than KI, it shows the thinning of the outer layers
more drastically. Finally the CaI 657.3\,nm seems not to change.
As the blue spectral range was not covered in 2002 the other low
excitation species cannot be compared.

The ratio of the CaI 657.3\,nm to CaI 422.7\,nm gives, for an
optically thin line, an electron temperature of 1.000$\pm$30\,K.
This line ratio is very sensitive to temperature. Both lines are
getting optically thick very quickly due to the high abundance of
the element. Thus it is likely that they originate from a thin,
and thus isothermal, layer at the outermost part.

Assuming solar abundances for NaI and 766.5\,nm KI I derive an
excitation temperature of 1.050\,K. The FeI lines has strong
variations in their transition probabilities. Therefore the
optically depths vary from line to line. We "see" different
regions with each line. No real temperatures can be derived. All
solutions tend to go towards temperatures above 1.700K. The
optically thin solutions for CrI 425.4, 427.5, 429.0 + 520.6\,nm
(in fact a blend of 520.45 + 520.6 + 520.8\,nm) gives 1.350\,K.
But again the line with the highest transition probability
(429.0\,nm) is overestimated in the calculation. This again leads
to line self absorption and thus underestimates the real
temperature. This is especially supported by the observed line
strength of the 520.6\,nm feature. This only identified transition
in my spectrum not going to the ground state of the atom. On the
other hand it has an unknown FeI contamination. The RbI doublet
seems to be contaminated (see discussion above). For all other
elements (MnI, MgI, SrI) only single lines are available. Thus
temperatures can be derived only by assuming an abundance,
relative to other elements. Using solar abundance temperatures
like those for FeI are derived.

\section{Results}

V4332 Sgr underwent a mysterious eruption. The nature of the event
remains unclear. The photometric behavior before the outburst with
an amplitude of $\Delta R \ge 2\fm4$ is an argument against the
scenario of a main sequence stellar merger (Soker \& Tylenda 2003,
Tylenda et al. 2004). Such a high amplitude cannot be produced by
an eclipsing MS binary system. Also the merger should give a
sudden onset of the envelope expansion. Thus the brightening
cannot be a forerunner of the event. This applies even more for
the planet destruction suggested by Retter \& Marom
\cite{planets}. This amplitude resembles more likely the
variations of CVs from "on" to "off" phases. The stable red band
photometry (the bolometric correction for the $R$ band is for such
stars rather small) and the (marginal) luminosity increase during
the recent years also is not expected for a post-merger scenario.
Tylenda et al. (2004) predict a steady decreasing luminosity
($\approx 1\fm0$ since the HST data). This clearly is not the
case. The star has not declined completely to its pre-outburst
luminosity yet, and shows a much cooler atmosphere than that of
the pre-outburst.

The photometry and the investigation of the underlying continuum
reveals a K8-M0 stellar atmosphere. This leads to a revision to
earlier studies, that the circumstellar dust, built 1998-2000, do
not give any continuum extinction. For such a photosphere the
luminosity is fitted to the photometric data by

\begin{equation}
\log L = 2 \times \log D - 5.7063
\end{equation}

\noindent where $D$ is the distance in units of $pc$. At the distances derived
from the progenitor this luminosity is too high for am M0V star.

Dust (assuming no shock heating) at a temperature of 865\,K is
then at a distance {of about 10} stellar radii ($R_*$). As there
are found no evidence for fast flows (e.g. no HeI 1.08$\mu$m and
no P-Cygni line profiles) this assumption seems do be reasonable.
The dust gets rapidly cooler at larger distances. Thus already at
a distance of $13-15 R_*$, even assuming constant density, the
contribution to the NIR excess gets negligible. Thus a dust mass
estimate from the NIR flux leads to an underestimate of the real
values. Estimates using the NILFISC code (Koller \& Kimeswenger
2001) for the circumstellar extinction leads to a much larger
extent of the absorbing dust shield. Without a detailed
geometrical model (disk vs. shell and density profile) this
question cannot be settled. Some hints for a complex non spherical
geometry might be given by the fact that the atomic lines are
blueshifted with respect to the molecular lines. But higher
spectroscopic resolution is needed to confirm this. An observation
like it was tried recently for V838 Mon by Lane et al. (2005) up
to now fails due to the faintness of the target even in NIR bands.

For a complete analysis of the emission spectrum a full radiative
transfer (RT) have to be calculated. But as the current
observations do not allow to verify whether the emitting layers
have a hydrostatic stratification or are part of an self
contracting shell. As the line optical depth is clearly above
unity in the lines and, as the underlying stellar photosphere
gives optical depths below unity in the continuum, a full non-LTE
radiative transfer have to be calculated to derive a full spectrum
and reliable  abundances. All this requires some assumptions on
the geometry.

For the investigation of the geometry, very high spectral resolution giving
line profiles and the relative blue/redshifts of the lines are required.

The cold material is either a shell of very clumpy clouds, or -
more likely - an equatorial disk. The latter makes, so massive
shortly after the event, a scenario with a single star (e.g. an He
shell flash) very unlikely. A Nova event with a massive shell, as
suggested by Martini et al. \cite{martini99} to model the moving
"quasi photosphere", might be a possible explanation. Shara et al.
\cite{nova1} and Prialnik \& Kovetz \cite{nova2} predict such
massive shells on WDs with masses as low as 0.6 M$_\odot$ - thus
novae during their first few recurrent events. This is not in
contradiction with the missing of $^{26}$Al (Banerjee et al.
2004b). Starrfield et al. \cite{nova3} points out that the ratio
$^{26}$Al/$^{27}$Al in Novae with low mass WDs might be as low as
0.01. Also the low expansion velocities measured for V4332 Sgr
(Martini et al. 1999) are in agreement with the low mass WD models
(Kato \& Iben 1992). Even if the model of Iben \& Tutukov
\cite{ibtu} does not work directly, the results of Shara et al.
\cite{nova1} for a 0.4\,M$_\odot$ He-WD look promising for the
multiple outbursts like those seen in V838 Mon and should be
discussed also with respect to this kind of objects. They show in
their model a delay of the second outburst when waiting for the
expansion of the envelope after convection recedes from the
surface.

V838 Mon and V4332 Sgr are sometimes also discussed in relation to
born-again PNe and late He-flash post-AGB scenarios. The ScO and
AlO molecules in emission are a signature of a very O-rich ejecta.
This is somehow in contradiction to the models, predicting a
C-rich shell (Herwig et al. 1999).

Generally  it is somewhat puzzling to me that the disk seemed to
be formed years later after outburst when the photometric behavior
have already "stabilized". The pre-outburst behavior, rising
during several decades, differs completely from that of V838 Mon.
That object was stable before the outburst (Goranskij et al.
2004). Boschi \& Munari \cite{bomu} suggest in their conclusion
that {\it "this kind of results support the need of radically new
models for M31-RV and V838 Mon and V4332 Sgr."}

\section*{Acknowledgments}
I like to thank P. Martini for pointing out my missing of the withdrawal of the
1968 outburst by Sharov in my paper on V838 Mon and for the CCD images for the
calibration of the surrounding CCD sequence of the object. Thanks also to D.
Fr\"obich for sending me his
extinction charts in digital form.\\
This research has made use of the SIMBAD database, operated at CDS
(Strasbourg, France), the SuperCOSMOS Sky Surveys at the WFAU/ROE
(Edinburgh, UK), the USNO Integrated Image Archive Service
(Flagstaff, USA) and of the Asiago Observatory Archive (Padova,
I).

\end{document}